\documentclass[11pt,a4paper]{article}
\usepackage[hyperref]{emnlp2018}
\usepackage{times}
\usepackage{latexsym}
\usepackage{algorithm}

\usepackage{graphicx}
\usepackage{subfigure}
\usepackage{url}
\usepackage[noend]{algpseudocode}
\usepackage{algorithmicx,algorithm}
\usepackage{xcolor}
\usepackage{amsmath}
\aclfinalcopy 
\def\aclpaperid{43} 

\title{Attention-Based  Capsule Networks  with Dynamic Routing  \\ for Relation Extraction}
\author{Ningyu Zhang\textsuperscript{1}   \quad
Shumin Deng\textsuperscript{2,3}    \quad
 Zhanlin Sun\textsuperscript{2}  \quad
 Xi Chen\textsuperscript{2}  \\
  \textbf{Wei Zhang\textsuperscript{3,4}} \quad
   \textbf{Huajun Chen\textsuperscript{2}}\thanks{\quad Corresponding author.} \\
 1. Artificial Intelligence  Research Institute, Zhejiang Lab, China \\
 2. College of Computer Science and Technology, Zhejiang University, China \\
 3. Alibaba-Zhejiang University Frontier Technology Research Center, China \\
 4. Alibaba Group, China\\
  zhangningyu@zju.edu.cn }

\date{}

\begin{document}
\maketitle
\begin{abstract}
A capsule is a group of neurons, whose activity vector represents the instantiation parameters of a specific type of entity.
In this paper, we explore the capsule networks used for relation extraction in a multi-instance multi-label learning framework and propose a novel neural approach based on   capsule networks with attention mechanisms. We evaluate our method with different benchmarks, and it is demonstrated that our method improves the precision of the predicted relations. Particularly, we show that capsule networks improve multiple entity pairs relation extraction\footnote{In this paper, multiple entity pairs relation extraction refers to multiple entity pairs  in a single sentence and  each pair of entities  contains only one relation label.}.

\end{abstract}

\section{Introduction}
This paper focus on the task of relation extraction. One popular method for relation extraction is the multi-instance multi-label learning framework (MIML) \cite{surdeanu2012multi} with distant supervision, where the mentions for an entity pair are aligned with the relations in Freebase \cite{bollacker2008freebase}. The recently proposed  neural network (NN) models \cite{zeng2014relation,ye2017jointly,yang2018ensemble,wangguanying} achieve state-of-the-art performance.  However, despite the great success of these NNs, some disadvantages remain. First, the existing models focus on, and heavily rely on, the quality of instance representation. Using a vector to represent a sentence is limited  because languages are delicate and complex. Second, CNN subsampling fails to retain the precise spatial relationships between higher-level parts. The structural relationships such as the positions in sentences are valuable.  Besides,  existing   aggregation operations summarizing the sentence meaning into a fixed-size vector such as max or average pooling  are lack of guidance by task information. Self-attention \cite{lin2017structured} can select task-dependent information. However, the context vectors are fixed once learned \cite{gong2018information}.

More importantly, most state-of-the-art systems can only predict one most likely relation for a single entity pair. However, it is very common that one sentence may contain multiple entity pairs and describe multiple relations. It is beneficial to consider other relations in the context while predicting the  relations \cite{sorokin2017context}.
For example, given the sentence "\emph{[\textbf{Swag It Out}] is the official  debut  single  by American  [\textbf{singer}] [\textbf{Zendaya}]}", our model can predict multi-relations for these multiple entity pairs simultaneously. 



In our work, we  present a novel architecture based on  capsule networks \cite{sabour2017dynamic} for  relation extraction. We regard the aggregation as a routing problem of how to deliver the messages from source nodes to target nodes.  This process enables the capsule networks to decide what and how much information need to be transferred  as well as  identify complex and interleaved features.  Furthermore, the capsule networks convert  the  multi-label classification problem into a multiple binary classification problem.  
We also import word-level  attention  by considering the different contribution of  the words.  The experimental results show that our model achieves improvements in both single and multiple relation extraction.

\begin{figure*}
\centering
\includegraphics [width=1.0\textwidth]{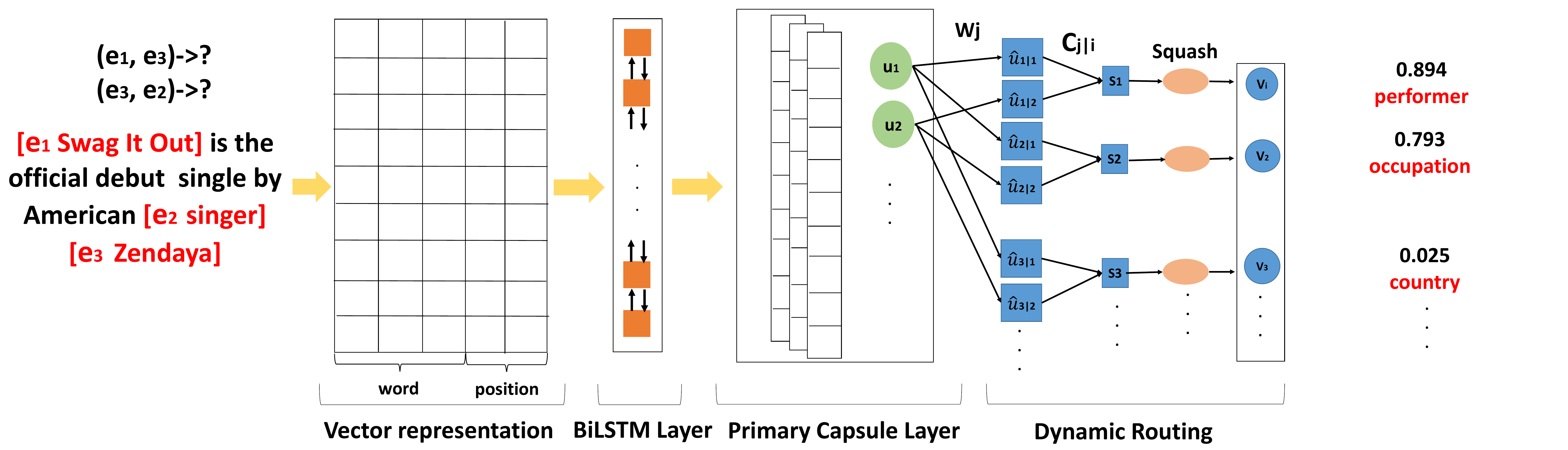}
  \caption{Architecture of capsule networks for relation extraction}
  \label{fig: architecture}
\end{figure*}

\section{Related Work}
\textbf{Neural Relation Extraction:} In the recent years, NN models have shown superior performance over approaches using hand-crafted features in various tasks.
CNN is the first one of the deep learning models that have been applied to relation extraction \cite{santos2015classifying}. Variants of convolutional networks include piecewise-CNN (PCNN) \cite{zeng2015distant}, instance-level selective attention CNN \cite{lin2016neural},        rank CNN \cite{ye2017jointly}, attention and memory CNN \cite{feng2017effective} and syntax-aware CNN \cite{he2018see}. Recurrent neural networks (RNN) are another popular choice, and have been used in recent works in the form of  attention RNNs \cite{zhou2016attention}, context-aware long short-term memory units (LSTMs) \cite{sorokin2017context}, graph-LSTMs \cite{peng2017cross} and ensemble LSTMs \cite{yang2018ensemble}.

\textbf{Capsule Network:}
Recently, the capsule network has been proposed to improve the representation limitations of CNNs and RNNs. \cite{sabour2017dynamic} replaced the scalar-output feature detectors of CNNs with vector-output capsules and max-pooling with routing-by-agreement. \cite{hinton2018matrix}) proposed a new iterative routing procedure among capsule layers, based on the EM algorithm. For natural language processing tasks, \cite{zhao2018investigating} explored capsule networks for text classification. \cite{gong2018information} designed two dynamic routing policies to aggregate the outputs of RNN/CNN encoding layer into a final encoding vector.  \cite{wang2018sentiment}  proposed a capsule model based on RNN for sentiment analysis. To the best of our knowledge, there has been no work that investigates the performance of capsule networks in relation extraction tasks at present.

\section{Methodology}
In this section, we first introduce the MIML framework, and then describe the model architecture we propose for relation extraction, which is shown in Figure \ref{fig: architecture}.

\subsection{Preliminaries}
In MIML, the set of text sentences for the single entity pair or multiple entity pairs\footnote{Since the number of sentences with multiple entity pairs is relatively small,  we make use of all the sentences as training samples.} (maximum two entity pairs in this paper) is denoted by $X = \{x_1,x_2, . . . ,x_n\}$.
Assumed that there are $E$ predefined relations (including NA) to extract. Formally, for each relation $r$, the prediction target is denoted by $P(r | x_1, . . . ,x_n)$.

\subsection{Neural Architectures}

\textbf{Input Representation:} For each sentence $x_i$, we use pretrained word embeddings to project each word token onto the $d_w$-dimensional space.  We adopt the position features    as  the combinations of the relative distances from the current word to $M$ entities  and encode these distances in $M$ $d_p$-dimensional vectors\footnote{We also adopt an   attention  mechanism  similar to word-level attention in Bi-LSTM layer by considering the different contribution of  the $M$ position embeddings.}.  For single entity pair relation  extraction, $M = 2$; for multiple entity pairs relation extraction, we limit the maximum number of entities in a sentence to four (i.e. two entity pairs).  As three entities in one instance is possible  when two tuples have a common entity, we set the relative distance to the missing entity to a very large number.  Finally, each  sentence is transformed into a matrix $x_i=\{w_1,w_2,...,w_L\} \in R^{L\times V}$, where $L$ is the sentence length with padding and $V=d_w+d_p *M$.

\textbf{Bi-LSTM  Layer:}
We make use of LSTMs to deeply learn the semantic meaning of a sentence. We concatenate the current memory cell hidden state vector $h_t$ of LSTM from two directions as the output vector $h_t=[\overrightarrow h_t, \overleftarrow h_t] \in R^{2B}  $ at time $t$, where $B$ denotes the dimensionality of LSTM.

We import word-level attention mechanism as only a few words in a sentence that are relevant to the relation expressed \cite{jat2018improving}.
The scoring function is $g_t= h_t \times A \times r$, where $A \in R^{E\times E}$ is a square matrix and $r \in R^{E\times1}$ is a relation vector. Both $A$ and $r$ are learned.
After obtaining $g_t$, we feed them to a softmax function to calculate the final importance $\alpha_t=softmax(g_t)$. Then, we get the representation $\tilde x_t=\alpha_th_t$.

For a given bag of sentences, the learning is done using the setting proposed by \cite{zeng2015distant}, where the sentence with highest probability of expressing the  relation in a bag is selected to train the model in each iteration.

\textbf{Primary Capsule Layer:}
Suppose $u_i \in R^{d}$ denotes the instantiated parameters set of a capsule, where $d$ is the dimension of the capsule. Let $W^b \in R^{2 \times 2B}$ be the filter shared across different windows. We have a window sliding each 2-gram vector in the sequence
$\tilde x \in R^{L \times 2B}$ with stride 1 to produce a list of
capsules $U \in R^{(L+1)\times C \times d}$, totally $C \times d$ filters.
\begin{equation}
u_{ij} = squash(W^b_i \otimes S_{j-1:j} +b_1)
\end{equation}
where  $0\leq i\leq C \times d$,  $0\leq j\leq L+1$, $squash(x)=\frac{||x||^2}{0.5+||x||^2}\frac{x}{||x||}$, $b_1$ is the bias term.  For all $C \times d$ filters, the generated capsule feature maps can be rearranged as $U =\{u_1,u_2,...,u_{(L+1)\times C}\}$,  where totally $(L+1) \times C $ $d$-dimensional vectors are collected as capsules.

\begin{algorithm}[th]
\caption{Dynamic Routing Algorithm}
\label{algorithm: Dynamic Routing}
\begin{algorithmic}[1]
\State \textbf{procedure} ROUTING($\hat u_{j|i},\hat a_{j|i},r,l$)
\State Initialize the logits of coupling coefficients $b_{j|i}=0$
\For{$r$ iterations } 
　　\For{all capsule $i$ in layer $l$ and capsule $j$ in layer $l+1$}
　\State $c_{j|i}=\hat a_{j|i}\cdot softmax(b_{j|i})$
   \EndFor

\For{all capsule $j$ in layer $l+1$}
\State $v_j=squash(\sum_ic_{j|i}\hat u_{j|i}), a_j=||v_j||$
　　   \EndFor

\For{all capsule $i$ in layer $l$ and capsule $j$ in layer $l+1$}

\State $b_{j|i}=b_{j|i}+\hat u_{j|i}\cdot v_j$
　　   \EndFor
\EndFor
 \Return $v_j,a_j$
\end{algorithmic}
\end{algorithm}

\textbf{Dynamic Routing:}
We explore the transformation matrices to generate the prediction vector $u_{j|i} \in R^{d}$ from a child capsule $i$ to its parent capsule $j$. The transformation matrices share weights $W^c \in R^{E× \times d \times d}$  across the child capsules, where $E $ is the number relations (parent capsules) in the layer above. Formally, each corresponding vote can be computed by:
\begin{equation}
\hat u_{j|i}= W_j^cu_i+\hat b_{j|i} \in R^{d}
\end{equation}
The basic idea of dynamic routing is to design a nonlinear map:

$ \{\hat u_{j|i}\in R^{d}\}_{i=1,...,H, j=1,...,E}  \mapsto \{v_j\in R^{d}\}_{j=1}^E    $

where $H=(L+1) \times C$.

Inspired by \cite{zhao2018investigating}, we attempt to use the probability of existence of parent capsules  to iteratively amend the connection strength, which is summarized in Algorithm \ref{algorithm: Dynamic Routing}.  The length of the vector $v_j$ represents the  probability  of each relation. We use a separate margin loss  $L_k$  for each relation capsule $k$:
\begin{equation}\begin{split}
L_k&=Y_kmax(0,m^+-||v_k||)^2 \\
&+\lambda(1-Y_k)max(0,||v_k||-m^-)^2\end{split}
\end{equation}
where $Y_k = 1$ if the relation $k$ is present
, $m^+ = 0.9$ , $m^- = 0.1$ and $\lambda = 0.5$. The total loss can be formulated as: $L_{total} = \sum_{k=1}^{E}L_k$

\subsection{Prediction}
For single entity pair relation extraction,  we  calculate the length of the  vector $v_j$  which represents the  probability  of each relation. For multiple entity pairs relation extraction, we choose  relations with  top two  probability meanwhile  bigger than the threshold (We empirically set the threshold  0.7). Finally, we may get one or two predicted relations $r$.  Given  entity pair $(e_1,e_2)$, in order to choose which relationship the tuple  belongs to, we  adopt the pretrained embeddings of entities and relations\footnote{http://openke.thunlp.org}  and calculate $r_k = \mathop{\arg\min}\limits_{k}   |t-h-r_k|$ , where $t$, $h$ are the  embeddings  of entities  $e_1$,  $e_2$ respectively and $r_k$ is the  relation embedding.  The relation with  the closest embedding  to the entity embedding difference  is the predicted category.

\section{Experiments}
We test our  model on  the  NYT dataset (NYT) developed by \cite{riedel2010modeling}  for single entity pair relation extraction and the Wikidata dataset \cite{sorokin2017context} for multiple entity pairs relation extraction.  We exclude sentences longer than $L$ .  All code is implemented in Tensorflow \cite{abadi2016tensorflow}. We adopt the Adam optimizer   \cite{kingma2014adam} with learning rate 0.001, batch size 128, LSTMs' unit size 300,  $L=120$, $d_p = 5$, $d=8$, $C=32$,  dropout rate 0.5, routing iteration 3.

\subsection{Practical Performance}
\textbf{NYT dataset  (Single entity pair):}
\begin{figure}[htbp]
  \centering
  \includegraphics[width=0.48\textwidth]{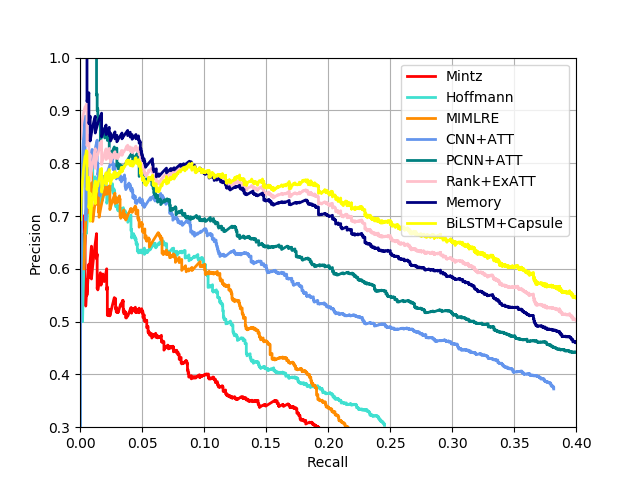}
  \caption{Performance comparison on the NYT dataset.}
  \label{fig:digit}
\end{figure}

\begin{figure}[htbp]
  \centering
  \includegraphics[width=0.49\textwidth]{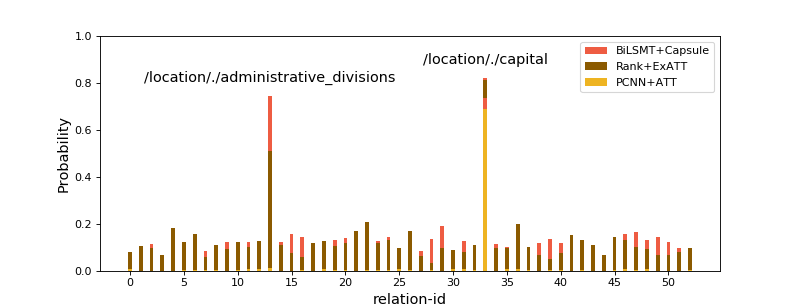}

  \caption{Normalized output relation scores.}
  \label{fig:digit1}
\end{figure}

We utilize the word embeddings released by \cite{lin2016neural}\footnote{$d_w = 50 $}.  The precision-recall curves for different models on the test set are shown in Figure \ref{fig:digit}.  Our model  BiLSTM+Capsule achieves comparable results compared with all baselines, where Mintz refers to \cite{mintz2009distant}, Hoffmann refers to \cite{hoffmann2011knowledge},  MIMLRE refers to \cite{surdeanu2012multi}, CNN+ATT refers to \cite{zeng2015distant}, PCNN+ATT refers to \cite{lin2016neural},  Rank+ExATT refers to \cite{ye2017jointly} and Memory refers to \cite{feng2017effective}.  We also show the precision numbers for some particular recalls as well as the AUC in Table \ref{tab: precisionsNYT}, where our model  generally leads to better precision. Interestingly, we observe our model  achieve  comparable results  to predict  multi-relation compared with Rank+ExATT in Figure \ref{fig:digit1}. Given an entity tuple (South Korea, Seoul) which has two relations: /location/./administrative\_divisions  and /location/./capital. We observe these two relations have the highest scores among the other relations in our model which demonstrate the ability of multi-relations prediction.

\begin{table}[!htbp]
\centering
\caption{Precisions on the NYT dataset.}
\label{tab: precisionsNYT}
\begin{small}
\begin{tabular}{p{1.7cm}|p{0.65cm}|p{0.65cm}|p{0.65cm}|p{0.65cm}|p{0.65cm}}
\hline
\hline
\centering Recall&\centering 0.1&\centering 0.2&\centering 0.3&\centering 0.4&AUC\\
\hline
\centering PCNN+ATT&0.698&0.606&0.518&0.446   &0.323\\
\hline
\centering Rank+ExATT &0.789&0.726&0.620&0.514  &0.395\\
\hline
\centering Our Model&0.788&\textbf{0.743}&\textbf{0.654}&\textbf{0.546}& \textbf{0.397}\\

\hline
\hline
\end{tabular}
\end{small}
\end{table}

\textbf{Wikidata dataset (Multiple entity pairs):}

We train  word embeddings  using Glove \cite{pennington2014glove}\footnote{$d_w = 200$} on the Wikipedia Corpus.   We show the precision numbers for some particular recalls as well as the AUC in Table \ref{tab:precisionsWiki},  where PCNN+ATT (1) refers to train sentences with two entities and one relation label,
PCNN+ATT (m) refers to train sentences  with four entities\footnote{Two additional position embeddings.} and two relation labels. 
We observe that our model exhibits the best performances.
 Moreover, in the process of predicting the existence of  relations for a sentence, our approach is more convenient , as the PCNN-ATT (1) has to predict all possible pairs of entities in the sentence while our approach can predict multiple relations simultaneously.

\begin{table}[!htbp]
\centering
\caption{Precisions on the Wikidata dataset.}
\label{tab:precisionsWiki}
\begin{small}
\begin{tabular}{c|c|c|c|c}
\hline
\hline
\centering Recall& \centering 0.1& \centering 0.2& \centering 0.3&   AUC\\

\hline

\centering Rank+ExATT & \centering 0.584& \centering0.535& \centering0.487& 0.392  \\
\hline
\centering PCNN+ATT (m)&0.365&0.317&0.213  &0.204\\

\hline
\centering PCNN+ATT (1)&0.665&0.517&0.413   &0.396\\
\hline
\centering Our Model&0.650&0.519&0.422& \textbf{0.405}\\

\hline
\hline
\end{tabular}
\end{small}
\end{table}
\textbf{Ablation study:}
To better demonstrate the  performance of  capsule net and attention mechanism, we remove the primary capsule layer and dynamic routing  to  make Bi-LSTM layer followed by  a fully connected layer  instead.  We also remove the word-level attention  separately.  The  experimental results on Wikidata dataset are summarized in Table \ref{tab:ablationwiki}. The results of "-Word-ATT"  row  refers to the results without  word-level attention. According to the table,  the drop of precision demonstrates that the  word-level attention is quite useful.  Generally, all two proposed  strategies contribute to the effectiveness of our model.

\begin{table}[!htbp]
\centering
\caption{Ablation study of capsule net and word-level attention  on Wikidata dataset.}\label{tab:ablationwiki}
\begin{tabular}{p{2.3cm}|p{0.7cm}|p{0.7cm}|p{0.7cm}|p{0.7cm}}

\hline
 \centering Recall& \centering 0.1& \centering 0.2& \centering 0.3&   AUC\\

\hline

\centering -Word-ATT    &0.648&0.515&0.395 &0.389\\

\hline
\centering -Capsule     &0.635&0.507&0.413   &0.386\\
\hline
\centering Our Model&0.650&0.519&0.422& 0.405\\

\hline

\end{tabular}

\end{table}

\subsection{Discussion}
\textbf{CNN vs  Capsule:}
Capsule networks achieve  comparable results compared with baselines. In fact, the capsule  combines features by clustering. A nonlinear map is constructed in an iterative manner, ensuring the output of each capsule to be sent to an appropriate parent in the subsequent layer. Dynamic routing may be more effective than the strategies such as max-pooling in CNN, which essentially detects whether a feature is present in any position of the text or not, but loses spatial information of the feature. Additionally, capsule achieves comparable results to predict multi-relations in the case of single entity pair, and performs better in the case of multiple entity pairs relation extraction.

\textbf{Choice of $d$}:
In the experiments, the larger the dimension of the capsule, the more the capabilities of the feature information it contains. However, larger dimension increases the computational complexity.
We test different levels of dimensions of capsules. The model is trained on two Nvidia  GTX1080ti GPUs
with 64G RAM and six Intel(R) Core(TM) i7-6850K CPU  3.60GHz.  As the table \ref{tab:choicedc} depicts,  the training time increases  with  the growth of $d$. When $d=32$, we observe  that the loss  decreases very  slowly and the model is difficult to converge.  So we only train 2 epochs and stop training.
We set the parameter $d=8$  empirically to balance the precision and training time cost.

\begin{table}[!htbp]
\centering
\caption{Precisions on the  Wikidata dataset with different  choice of $d$.}\label{tab:choicedc}
\begin{tabular}{p{1.5cm}|p{0.7cm}|p{0.7cm}|p{0.7cm}|p{0.7cm}|p{0.7cm}}

\hline
 \centering Recall& \centering 0.1& \centering 0.2& \centering 0.3&  AUC&Time   \\

\hline

\centering$d=1$&  0.602&0.487& 0.403& 0.367 &4h\\
\hline
\centering$d=32$  &0.645&0.501&0.393 &0.370&-\\

\hline
\centering   $d=16$  &0.655&0.518&0.413   &0.413& 20h\\
\hline
\centering $d=8$&0.650&0.519&0.422& 0.405& 8h\\

\hline

\end{tabular}

\end{table}

\textbf{Effects of Iterative Routing:}

\begin{table}[!htbp]
\centering
\caption{Precisions on the  Wikidata  dataset with different  number of dynamic routing iterations.}\label{tab:iterations}
\begin{tabular}{p{2.3cm}|p{0.7cm}|p{0.7cm}|p{0.7cm}|p{0.7cm}}

\hline
 \centering Recall& \centering 0.1& \centering 0.2& \centering 0.3&   AUC\\

\hline

\centering Iteration=1&  0.531&0.455& 0.353& 0.201 \\
\hline
\centering Iteration=2&0.592&0.498&0.385 &0.375\\

\hline
\centering Iteration=3&0.650&0.519&0.422& 0.405\\
\hline
\centering Iteration=4&0.601&0.505&0.422& 0.385\\
\hline
\centering Iteration=5&0.575&0.495&0.394& 0.376\\
\hline

\end{tabular}

\end{table}
We also study how the iteration number affect the performance on the Wikidata dataset. Table \ref{tab:iterations} shows the comparison of 1 - 5 iterations.  We find that the performance  reach the best when iteration is set to 3. The results indicate the dynamic routing is contributing to improve the performance. Specifically, in the iteration algorithm, the $b_{j|i}=b_{j|i}+\hat u_{j|i}  \cdot v_j$. When the number of iteration is very large, $v_j$ becomes either $0$ or $1$, which means each underlying capsule is only linked to a single upper capsule. Therefore, the iteration times should not be too large.

\subsection{Conclusion}

We propose a  relation extraction approach  based on  capsule networks with  attention mechanism. Although we use Bi-LSTM  as sentence encoding in this paper, the other encoding method, such as convolved n-gram, could be alternatively used.  Experimental results of two  benchmarks show that the model improves the precision of the predicted relations.

In the future,  we tend to resolve the situation of  how to assign  predicted relationship to multi entity pairs when two entities  have multi-relations   by utilizing prior knowledge such as entity type and  joint training with named entity recognition.  We will also try to optimize the model in terms of speed and  focus on other problems by leveraging class ties between labels, specially on multi-label learning problems. Besides, dynamic routing could also be useful to improve other  natural language processing  tasks such as   the sequence-to-sequence task and so on.

\section*{Acknowledgments}
We  want to express gratitude to the anonymous reviewers for their hard work and kind comments, which will further improve our work in the future. This work is funded by NSFC 61673338/61473260, and partly supported by Alibaba-Zhejiang University Joint Institute of Frontier Technologies.

\bibliography{emnlp2018}

\begin{thebibliography}{27}
\expandafter\ifx\csname natexlab\endcsname\relax\def\natexlab#1{#1}\fi

\bibitem[{Abadi et~al.(2016)Abadi, Barham, Chen, Chen, Davis, Dean, Devin,
  Ghemawat, Irving, Isard et~al.}]{abadi2016tensorflow}
Mart{\'\i}n Abadi, Paul Barham, Jianmin Chen, Zhifeng Chen, Andy Davis, Jeffrey
  Dean, Matthieu Devin, Sanjay Ghemawat, Geoffrey Irving, Michael Isard, et~al.
  2016.
\newblock Tensorflow: A system for large-scale machine learning.
\newblock In \emph{OSDI}, volume~16, pages 265--283.

\bibitem[{Bollacker et~al.(2008)Bollacker, Evans, Paritosh, Sturge, and
  Taylor}]{bollacker2008freebase}
Kurt Bollacker, Colin Evans, Praveen Paritosh, Tim Sturge, and Jamie Taylor.
  2008.
\newblock Freebase: a collaboratively created graph database for structuring
  human knowledge.
\newblock In \emph{Proceedings of the 2008 ACM SIGMOD international conference
  on Management of data}, pages 1247--1250. AcM.

\bibitem[{Feng et~al.(2017)Feng, Guo, Qin, Liu, and Liu}]{feng2017effective}
Xiaocheng Feng, Jiang Guo, Bing Qin, Ting Liu, and Yongjie Liu. 2017.
\newblock Effective deep memory networks for distant supervised relation
  extraction.
\newblock In \emph{Proceedings of the Twenty-Sixth International Joint
  Conference on Artificial Intelligence, IJCAI}, pages 19--25.

\bibitem[{Gong et~al.(2018)Gong, Qiu, Wang, and Huang}]{gong2018information}
Jingjing Gong, Xipeng Qiu, Shaojing Wang, and Xuanjing Huang. 2018.
\newblock Information aggregation via dynamic routing for sequence encoding.
\newblock \emph{arXiv preprint arXiv:1806.01501}.

\bibitem[{He et~al.(2018)He, Chen, Li, Zhang, Zhang, and Zhang}]{he2018see}
Zhengqiu He, Wenliang Chen, Zhenghua Li, Meishan Zhang, Wei Zhang, and Min
  Zhang. 2018.
\newblock See: Syntax-aware entity embedding for neural relation extraction.
\newblock \emph{arXiv preprint arXiv:1801.03603}.

\bibitem[{Hinton et~al.(2018)Hinton, Frosst, and Sabour}]{hinton2018matrix}
Geoffrey Hinton, Nicholas Frosst, and Sara Sabour. 2018.
\newblock Matrix capsules with em routing.

\bibitem[{Hoffmann et~al.(2011)Hoffmann, Zhang, Ling, Zettlemoyer, and
  Weld}]{hoffmann2011knowledge}
Raphael Hoffmann, Congle Zhang, Xiao Ling, Luke Zettlemoyer, and Daniel~S Weld.
  2011.
\newblock Knowledge-based weak supervision for information extraction of
  overlapping relations.
\newblock In \emph{Proceedings of the 49th Annual Meeting of the Association
  for Computational Linguistics: Human Language Technologies-Volume 1}, pages
  541--550. Association for Computational Linguistics.

\bibitem[{Jat et~al.(2018)Jat, Khandelwal, and Talukdar}]{jat2018improving}
Sharmistha Jat, Siddhesh Khandelwal, and Partha Talukdar. 2018.
\newblock Improving distantly supervised relation extraction using word and
  entity based attention.
\newblock \emph{arXiv preprint arXiv:1804.06987}.

\bibitem[{Kingma and Ba(2014)}]{kingma2014adam}
Diederik~P Kingma and Jimmy Ba. 2014.
\newblock Adam: A method for stochastic optimization.
\newblock \emph{arXiv preprint arXiv:1412.6980}.

\bibitem[{Lin et~al.(2016)Lin, Shen, Liu, Luan, and Sun}]{lin2016neural}
Yankai Lin, Shiqi Shen, Zhiyuan Liu, Huanbo Luan, and Maosong Sun. 2016.
\newblock Neural relation extraction with selective attention over instances.
\newblock In \emph{Proceedings of the 54th Annual Meeting of the Association
  for Computational Linguistics (Volume 1: Long Papers)}, volume~1, pages
  2124--2133.

\bibitem[{Lin et~al.(2017)Lin, Feng, Santos, Yu, Xiang, Zhou, and
  Bengio}]{lin2017structured}
Zhouhan Lin, Minwei Feng, Cicero Nogueira~dos Santos, Mo~Yu, Bing Xiang, Bowen
  Zhou, and Yoshua Bengio. 2017.
\newblock A structured self-attentive sentence embedding.
\newblock \emph{arXiv preprint arXiv:1703.03130}.

\bibitem[{Mintz et~al.(2009)Mintz, Bills, Snow, and
  Jurafsky}]{mintz2009distant}
Mike Mintz, Steven Bills, Rion Snow, and Dan Jurafsky. 2009.
\newblock Distant supervision for relation extraction without labeled data.
\newblock In \emph{Proceedings of the Joint Conference of the 47th Annual
  Meeting of the ACL and the 4th International Joint Conference on Natural
  Language Processing of the AFNLP: Volume 2-Volume 2}, pages 1003--1011.
  Association for Computational Linguistics.

\bibitem[{Peng et~al.(2017)Peng, Poon, Quirk, Toutanova, and
  Yih}]{peng2017cross}
Nanyun Peng, Hoifung Poon, Chris Quirk, Kristina Toutanova, and Wen-tau Yih.
  2017.
\newblock Cross-sentence n-ary relation extraction with graph lstms.
\newblock \emph{arXiv preprint arXiv:1708.03743}.

\bibitem[{Pennington et~al.(2014)Pennington, Socher, and
  Manning}]{pennington2014glove}
Jeffrey Pennington, Richard Socher, and Christopher Manning. 2014.
\newblock Glove: Global vectors for word representation.
\newblock In \emph{Proceedings of the 2014 conference on empirical methods in
  natural language processing (EMNLP)}, pages 1532--1543.

\bibitem[{Riedel et~al.(2010)Riedel, Yao, and McCallum}]{riedel2010modeling}
Sebastian Riedel, Limin Yao, and Andrew McCallum. 2010.
\newblock Modeling relations and their mentions without labeled text.
\newblock In \emph{Joint European Conference on Machine Learning and Knowledge
  Discovery in Databases}, pages 148--163. Springer.

\bibitem[{Sabour et~al.(2017)Sabour, Frosst, and Hinton}]{sabour2017dynamic}
Sara Sabour, Nicholas Frosst, and Geoffrey~E Hinton. 2017.
\newblock Dynamic routing between capsules.
\newblock In \emph{Advances in Neural Information Processing Systems}, pages
  3859--3869.

\bibitem[{Santos et~al.(2015)Santos, Xiang, and Zhou}]{santos2015classifying}
Cicero Nogueira~dos Santos, Bing Xiang, and Bowen Zhou. 2015.
\newblock Classifying relations by ranking with convolutional neural networks.
\newblock \emph{arXiv preprint arXiv:1504.06580}.

\bibitem[{Sorokin and Gurevych(2017)}]{sorokin2017context}
Daniil Sorokin and Iryna Gurevych. 2017.
\newblock Context-aware representations for knowledge base relation extraction.
\newblock In \emph{Proceedings of the 2017 Conference on Empirical Methods in
  Natural Language Processing}, pages 1784--1789.

\bibitem[{Surdeanu et~al.(2012)Surdeanu, Tibshirani, Nallapati, and
  Manning}]{surdeanu2012multi}
Mihai Surdeanu, Julie Tibshirani, Ramesh Nallapati, and Christopher~D Manning.
  2012.
\newblock Multi-instance multi-label learning for relation extraction.
\newblock In \emph{Proceedings of the 2012 joint conference on empirical
  methods in natural language processing and computational natural language
  learning}, pages 455--465. Association for Computational Linguistics.

\bibitem[{Wang et~al.(2018{\natexlab{a}})Wang, Wang, and Chen}]{wangguanying}
Guanying Wang, Ruoxu Wang, and Huajun Chen. 2018{\natexlab{a}}.
\newblock Label-free distant supervision for relation extraction via knowledge
  graph embedding.
\newblock In \emph{Proceedings of the 2018 Conference on Empirical Methods in
  Natural Language Processing}.

\bibitem[{Wang et~al.(2018{\natexlab{b}})Wang, Sun, Han, Liu, and
  Zhu}]{wang2018sentiment}
Yequan Wang, Aixin Sun, Jialong Han, Ying Liu, and Xiaoyan Zhu.
  2018{\natexlab{b}}.
\newblock Sentiment analysis by capsules.
\newblock In \emph{Proceedings of the 2018 World Wide Web Conference on World
  Wide Web}, pages 1165--1174. International World Wide Web Conferences
  Steering Committee.

\bibitem[{Yang et~al.(2018)Yang, Wang, and Li}]{yang2018ensemble}
Dongdong Yang, Senzhang Wang, and Zhoujun Li. 2018.
\newblock Ensemble neural relation extraction with adaptive boosting.
\newblock \emph{arXiv preprint arXiv:1801.09334}.

\bibitem[{Ye et~al.(2017)Ye, Chao, Luo, and Li}]{ye2017jointly}
Hai Ye, Wenhan Chao, Zhunchen Luo, and Zhoujun Li. 2017.
\newblock Jointly extracting relations with class ties via effective deep
  ranking.
\newblock In \emph{Proceedings of the 55th Annual Meeting of the Association
  for Computational Linguistics (Volume 1: Long Papers)}, volume~1, pages
  1810--1820.

\bibitem[{Zeng et~al.(2015)Zeng, Liu, Chen, and Zhao}]{zeng2015distant}
Daojian Zeng, Kang Liu, Yubo Chen, and Jun Zhao. 2015.
\newblock Distant supervision for relation extraction via piecewise
  convolutional neural networks.
\newblock In \emph{Proceedings of the 2015 Conference on Empirical Methods in
  Natural Language Processing}, pages 1753--1762.

\bibitem[{Zeng et~al.(2014)Zeng, Liu, Lai, Zhou, and Zhao}]{zeng2014relation}
Daojian Zeng, Kang Liu, Siwei Lai, Guangyou Zhou, and Jun Zhao. 2014.
\newblock Relation classification via convolutional deep neural network.
\newblock In \emph{Proceedings of COLING 2014, the 25th International
  Conference on Computational Linguistics: Technical Papers}, pages 2335--2344.

\bibitem[{Zhao et~al.(2018)Zhao, Ye, Yang, Lei, Zhang, and
  Zhao}]{zhao2018investigating}
Wei Zhao, Jianbo Ye, Min Yang, Zeyang Lei, Suofei Zhang, and Zhou Zhao. 2018.
\newblock Investigating capsule networks with dynamic routing for text
  classification.
\newblock \emph{arXiv preprint arXiv:1804.00538}.

\bibitem[{Zhou et~al.(2016)Zhou, Shi, Tian, Qi, Li, Hao, and
  Xu}]{zhou2016attention}
Peng Zhou, Wei Shi, Jun Tian, Zhenyu Qi, Bingchen Li, Hongwei Hao, and Bo~Xu.
  2016.
\newblock Attention-based bidirectional long short-term memory networks for
  relation classification.
\newblock In \emph{Proceedings of the 54th Annual Meeting of the Association
  for Computational Linguistics (Volume 2: Short Papers)}, volume~2, pages
  207--212.

\end{thebibliography}
\bibliographystyle{acl_natbib_nourl}

\end{document}